# Testing Geological Models with Terrestrial Antineutrino Flux Measurements

S.T. Dye [a,b,*]
[a] Hawaii Pacific University, Kaneohe, Hawaii, USA
[b] University of Hawaii, Honolulu, Hawaii, USA

* Corresponding author.
*E-mail address:* sdye@phys.hawaii.edu (S.T. Dye).

ABSTRACT
Uranium and thorium are the main heat producing elements in the earth. Their quantities and distributions, which specify the flux of detectable antineutrinos generated by the beta decay of their daughter isotopes, remain unmeasured. Geological models of the continental crust and the mantle predict different quantities and distributions of uranium and thorium. Many of these differences are resolvable with precision measurements of the terrestrial antineutrino flux. This precision depends on both statistical and systematic uncertainties. An unavoidable background of antineutrinos from nuclear reactors typically dominates the systematic uncertainty. This report explores in detail the capability of various operating and proposed geo-neutrino detectors for testing geological models.

## 1. Introduction

Detectable geo-neutrinos are electron antineutrinos from the decay series of uranium and thorium (Fiorentini, Lissia, and Mantovani, 2007). Large scintillating liquid detectors efficiently record their interactions, measuring their energy spectrum but not their direction (Araki et al., 2005). The predicted rate of geo-neutrino interactions depends strongly on proximity to continental crust (Mantovani et al., 2004). This is due to concentrations of uranium and thorium being enriched in this reservoir relative to those in the mantle and oceanic crust. Measurable variations in these concentrations exist among geological models, allowing potential discrimination by geo-neutrino observations. An unavoidable background to geo-neutrino observation comes from nuclear reactors, introducing irreducible measurement errors. This report compares the predicted geo-neutrino signals from various geological models for the continental crust and the mantle. It presents the calculated precision of background-subtracted geo-neutrino measurements by various detectors as a function of exposure. It concludes that significant potential exists for discriminating geological models by observations of geo-neutrinos at selected detection sites.

## 2. Geo-neutrino Detection

Geo-neutrino detection typically uses the same technology employed by reactor antineutrino experiments for many decades (Reines and Cowan, 1953). In this technique, a surface array of inward-looking photomultiplier tubes monitors a large central volume of scintillating liquid. Antineutrinos interact dominantly with free protons in the scintillating liquid. Correlated signals from the products of inverse neutron decay, $\bar{\nu}_e + p \rightarrow e^+ + n$, mark the interaction. Initially, a prompt positron provides a measure of the geo-neutrino energy. The subsequent neutron capture, depositing fixed energy, tags the event. This technique allows a spectral measurement of geo-neutrinos originating from the decay series of $^{238}$U and $^{232}$Th. Geo-neutrinos from all other isotopes, including $^{40}$K, lack the energy to initiate the inverse beta reaction on free protons. Figure 1 shows the calculated geo-neutrino energy spectrum. A project to directly measure these spectra is ongoing (Bellini et al., 2007). Note that the highest energy geo-neutrinos derive only from $^{238}$U. This enables separate measurement of geo-neutrinos from $^{238}$U and $^{232}$Th, offering an estimate of the integrated thorium to uranium ratio. The traditional inverse beta coincidence technique provides limited information on geo-neutrino direction (Apollonio et al., 1999). This impedes determination of geo-neutrino source locations and rejection of reactor background.

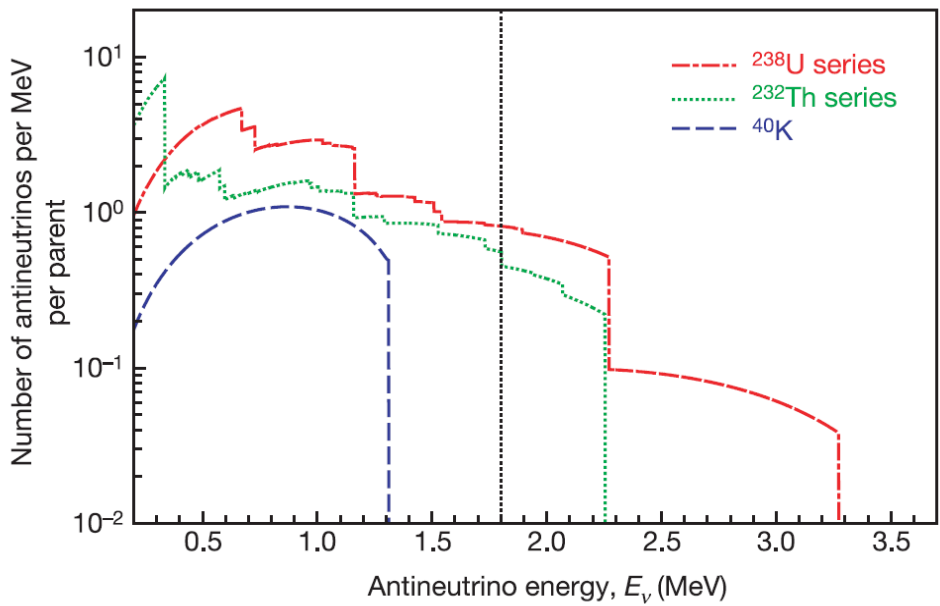

**Fig. 1.** Plot shows the energy spectra for antineutrinos from $^{40}$K decay and from the decay series of $^{238}$U and $^{232}$Th (Araki et al., 2005). The vertical line at 1.8 MeV (black dots) indicates the threshold energy for the inverse beta reaction.

## 3. Geo-neutrino Detectors

Large scintillating liquid instruments, typically deployed for physics and astrophysics investigations, detect geo-neutrino interactions. Their location and size critically determine the precision of geo-neutrino measurements. Two of the detectors, KamLAND at Kamioka, Japan (Abe et al., 2008), and Borexino at L'Aquila, Italy (Arpesella et al., 2008), are currently observing geo-neutrinos. A third detector, SNO+ at Sudbury, Canada, is under construction (Chen 2006). The remaining detectors in this study, DUSEL at Lead, South Dakota, Baksan in the Russian Caucasus (Barabanov et al., 2009), LENA at Pyhasalmi, Finland

(Hochmuth et al., 2006), and Hanohano at Hawaii (Learned et al., 2008), are proposed projects. Table I lists the location, latitude and longitude of these detection sites along with the size of the detectors and project status.

**Table 1**
Locations in N latitude and E longitude, project status, and sizes in $10^{32}$ free protons of large scintillating liquid detectors are presented.

|  | KamLAND | Borexino | SNO+ | DUSEL | Baksan | LENA | Hanohano |
|---|---|---|---|---|---|---|---|
| Location | Japan | Italy | Canada | S. Dakota | Russia | Finland | Hawaii |
| Latitude | 36.43 | 42.45 | 46.47 | 44.35 | 43.29 | 63.66 | 19.72 |
| Longitude | 137.31 | 13.57 | -81.20 | -103.75 | 42.70 | 26.05 | -156.32 |
| Status | Operating | Operating | Construction | Proposed | Proposed | Proposed | Proposed |
| Free $p^+$ | 0.62 | 0.18 | 0.57 | 36.7 | 4.0 | 36.7 | 7.34 |

## 4. Reference Signal and Background

The flux of antineutrinos at a given site is the sum of contributions from the crust, mantle, and commercial nuclear reactors. Contributions from other known sources, such as the diffuse supernova neutrino background and atmospheric neutrinos, are negligible in comparison. The flux varies significantly over the surface of the earth, being strongly influenced by the local crust and proximity to reactors. Integrating the product of flux, interaction cross section, and neutrino oscillation survival probability over the spectrum gives the event rate. Multiplying this rate by the detector exposure, given by the product of detector size and observation time, yields the expected number of antineutrino events observed. In practice detectors observe background events mimicking inverse neutron decay. This study ignores these background events, which originate from secondary cosmic rays and ambient radioactivity.

For comparing geological models of the continental crust and the mantle, this study employs a reference signal. It specifies the geo-neutrino detection rates from the crust and mantle by averaging predicted uranium and thorium concentrations from geological models (Mantovani et al., 2004; Fiorentini et al., 2007). Nuclear reactors contribute background in the geo-neutrino energy range (1.8 – 3.3 MeV) at a rate defined by their intensity and neutrino oscillation parameters (Abe et al., 2008). The calculated signal at each detector results from 201 reactors world-wide, operating continuously with a total power of 1.063 TW. Table 2 lists the reference signal rates from the crust and mantle and the background rate from reactors for each detection site. The reactor rate derives from the range of energy (3.3 – 9.0 MeV) greater than the maximum geo-neutrino energy. This is the measured quantity that estimates the reactor background rate in the geo-neutrino energy range. A factor calculated from the reactor spectrum provides the ratio of the rate in the geo-neutrino energy range to the rate in the greater energy range for each detection site. Figure 2 displays the spectra of reactor event rates at the detection sites. The wiggles in the spectra are due to neutrino oscillations (Abe et al., 2008), accounting for the differences in the calculated factors.

**Table 2**
Reference crust and mantle signals and high-energy reactor background rates in units of $(10^{32}p^+\text{-}y)^{-1}$ are listed for large scintillating liquid detectors. The contribution of oceanic crust to the crust signal is negligible at all listed sites except for Hanohano, where it is approximately 24%. The bottom entry is the calculated factor that estimates the reactor background rate in the geo-neutrino energy range from the high-energy reactor rate. Differences in these fractions are due to neutrino oscillations.

|  | KamLAND | Borexino | SNO+ | DUSEL | Baksan | LENA | Hanohano |
|---|---|---|---|---|---|---|---|
| crust, $c$ | 25.5 | 31.7 | 41.8 | 42.3 | 41.8 | 42.5 | 3.5 |
| mantle, $m$ | 9.0 | 9.0 | 9.0 | 9.0 | 9.0 | 9.0 | 9.0 |
| reactor, $r$ | 649.6 | 85.8 | 139.2 | 26.1 | 32.0 | 64.4 | 3.5 |
| $f$ | 0.318 | 0.347 | 0.314 | 0.345 | 0.344 | 0.345 | 0.348 |

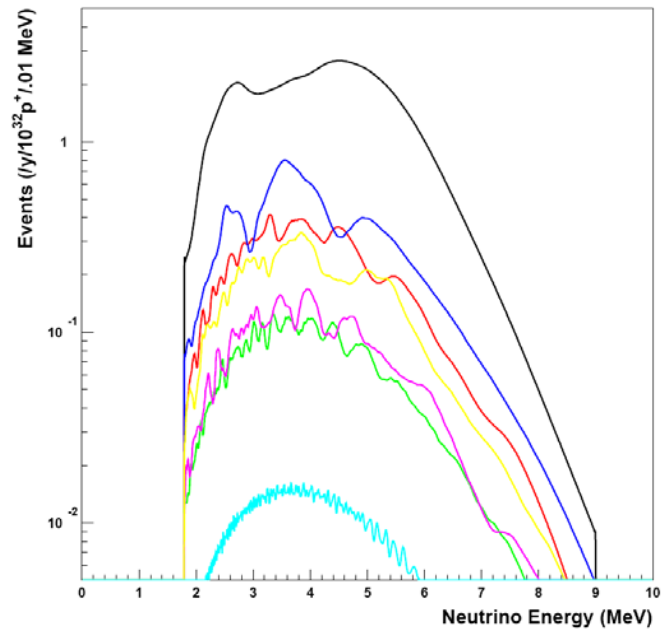

**Fig. 2.** Plot shows the different reactor antineutrino event rate energy spectra for the detection sites on a logarithmic scale. Black=KamLAND; Red=Borexino; Blue=SNO+; Green=DUSEL; Purple=Baksan; Yellow=LENA; Turquoise=Hanohano.

## 5. Geological Models

Geological models specifying different concentrations of uranium and thorium in crustal layers, and thereby different heat production rate densities (Rudnick and Fountain, 1995), predict different fluxes of geo-neutrinos from these reservoirs. This study assumes the physical structure of the crust as determined by seismology (Bassin et al., 2000) is common to all models, allowing the geo-neutrino flux to scale directly with radiogenic power. Table 3 compares the heat production rate density and radiogenic power for several models with the reference values. The models suggest powers of ~5 TW (Taylor and McLennan, 1985), ~8 TW (Weaver and Tarney, 1984; Rudnick and Fountain, 1995), and ~10 TW (Wedepohl, 1994; Shaw, 1986). This comparison suggests that a 10% measurement of the crustal geo-neutrino rate corresponding to one of these powers would exclude models predicting the other powers at the 99% CL (3σ) or better. It remains a challenge to resolve models predicting essentially the same power.

**Table 3**
Heat production rate densities for different geological models of the continental crust are compared with the reference. Heat production rate densities assume an average density of continental crust of 2.8 g cm$^{-3}$ (Rudnick and Fountain, 1995). Key to models: TM85 (Taylor and McLennan, 1985); WT84 (Weaver and Tarney, 1984); RF95 (Rudnick and Fountain, 1995); W94 (Wedepohl, 1994); S86 (Shaw, 1986).

|  | TM85 | WT84 | RF95 | Ref. | W94 | S86 |
|---|---|---|---|---|---|---|
| Heat production rate density (μW m$^{-3}$) | 0.58 | 0.92 | 0.93 | 1.03 | 1.25 | 1.31 |
| Power (TW) | 4.84 | 7.67 | 7.76 | 8.59 | 10.42 | 10.93 |
| % difference | -42 | -8 | -7 | 0 | +25 | +31 |

Geological models of the mantle predict different geo-neutrino fluxes from this reservoir. Calculation of fluxes employs a seismic model of the mantle (Dziewonski and Anderson, 1981) and assumes radial symmetry (Krauss, Glashow, and Schramm, 1984). Table 4 compares the radiogenic power and geo-neutrino rate specified by selected models with the reference values. Power and geo-neutrino rate do not scale directly due to the different radial distributions of uranium and thorium predicted by the various models. This comparison suggests that a 10% measurement of the mantle geo-neutrino rate corresponding to the prediction of one of the models would exclude all other models at the 95% CL (2σ) or better.

**Table 4**
Comparisons of the radiogenic power and geo-neutrino rate from the mantle as predicted by selected geological models. Key to models: TH05 (Tolstikhin and Hofmann, 2005); TKH06 (Tolstikhin, Kramers and Hofmann, 2006); KT97 (Kramers and Tolstikhin, 1997); TPW01 (Turcotte, Paul and White, 2001).

|  | TH05 | TKH06 | Ref. | KT97 | TPW01-I | TPW01-II |
|---|---|---|---|---|---|---|
| Power (TW) | 7.4 | 11.4 | 10.9 | 12.7 | 18.2 | 25.7 |
| Geo-nu (TNU) | 6.9 | 8.6 | 9.0 | 10.9 | 15.1 | 22.0 |
| %diff | -22 | -3 | 0 | +22 | +70 | +147 |

## 6. Uncertainties

The precision of the geo-neutrino rate measurements at a given observation site determines the capability for resolving geological models at that location. Both statistical and systematic uncertainties affect the precision. Statistical errors depend on the total antineutrino rate, reactor background rate, and exposure. Systematic errors depend on uncertainties in the detector exposure, antineutrino energy measurement, as well as signal and background rates. Measurements of geological interest include the total geo-neutrino rate, crust rate, mantle rate, and the thorium to uranium ratio. Estimating the ratio of events due to thorium to the events due to uranium requires dividing the geo-neutrino energy range at the endpoint of the thorium spectrum.

Neglecting events mimicking the inverse beta reaction, an observatory measures antineutrino events only. The total antineutrino rate in the geo-neutrino energy range (1.8 – 3.3 MeV) is the total number of recorded events divided by the exposure

$$n = N/\varepsilon = n_H + n_L = g + fr = c + m + fr,$$

where $n_H$ is the rate at higher energy (2.3 – 3.3 MeV: only uranium geo-neutrinos and reactor antineutrinos) and $n_L$ is the rate at lower energy (1.3 – 2.3 MeV: uranium plus thorium geo-neutrinos and reactor antineutrinos). The ratio of thorium geo-neutrinos to uranium geo-neutrinos is

$$\rho = f_U \left( \frac{n_H + n_L - fr}{n_H - frf_H} \right) - 1,$$

where $f_U = 0.54$ ($f_H = 0.84$) is the calculated fraction of uranium geo-neutrinos (reactor antineutrinos) in the higher energy range (2.3 – 3.3 MeV). Note that measurement of the geo-neutrino rates and the thorium to uranium ratio requires subtraction of background. For example, the crust rate is the total rate minus the calculated reactor rate plus the estimated mantle rate. Table 5 provides a key to the symbols representing quantities appearing in these equations.

**Table 5**
Key to symbols for quantities measured by a geo-neutrino observatory is below.

| | | | Rates | | | | | |
|---|---|---|---|---|---|---|---|---|
| Total | Low E | High E | Geo-nu | Crust | Mantle | Reactor | Th/U | Exposure |
| $n$ | $n_L$ | $n_H$ | $g$ | $c$ | $m$ | $r$ | $\rho$ | $\varepsilon$ |

Both statistical and systematic uncertainties degrade the precision with which an observatory measures the geo-neutrino rates and the thorium to uranium ratio. The following equations for the square of the uncertainties in the rates and thorium to uranium ratio are in terms of measured and calculated quantities. In the equations for the uncertainty in the geo-neutrino rates, the first term on the right side of the equations accounts for the statistical error and the following terms contribute to the systematic error. The statistical and systematic uncertainties in the thorium to uranium ratio are separate for ease of presentation.

$$\delta g^2 = (n + f^2 r)/\varepsilon + \sigma_f^2 (fr)^2 + \sigma_\varepsilon^2 (n - fr)^2$$

$$\delta c^2 = (n + f^2 r)/\varepsilon + \sigma_f^2 (fr)^2 + \sigma_m^2 m^2 + \sigma_\varepsilon^2 (n - fr - m)^2$$

$$\delta m^2 = (n + f^2 r)/\varepsilon + \sigma_f^2 (fr)^2 + \sigma_c^2 c^2 + \sigma_\varepsilon^2 (n - fr - c)^2$$

$$\delta \rho_{stat}^2 = \frac{f_U^2}{\varepsilon} \left\{ rn^2 \left[ \frac{f_H f}{(n_H - fr f_H)^2} \right]^2 + n_H \left[ \frac{fr(1 - f_H) - n_L}{(n_H - fr f_H)^2} \right]^2 + n_L \left[ \frac{1}{n_H - fr f_H} \right]^2 \right\}$$

$$\delta \rho_{sys}^2 = f_U^2 \left\{ \sigma_{f_U}^2 \left[ \frac{n - fr}{n_H - fr f_H} \right] + \sigma_f^2 \left[ \frac{fr(n f_H - n_H)}{(n_H - fr f_H)^2} \right]^2 + \sigma_{f_H}^2 \left[ \frac{fr f_H (n - fr)}{(n_H - fr f_H)^2} \right]^2 \right\}$$

Since the rate of reactor antineutrinos in the geo-neutrino energy region follows from the rate measured at energy higher than the geo-neutrino energy range, the uncertainty in the ratio of these rates stems from uncertainties in the absolute energy scale and neutrino oscillation parameters.

$$\sigma_f^2 = \sigma_e^2 + \sigma_o^2$$

Calculation of the measurement precisions requires values for the fractional uncertainties of the various quantities. Table 6 lists estimates for the systematic uncertainties.

**Table 6**
Systematic uncertainties for the quantities used to calculate measurement precision.

| Exposure | Crust rate | Mantle rate | Energy | Oscillation | Uranium | Reactor |
|---|---|---|---|---|---|---|
| $\sigma_\varepsilon$ | $\sigma_c$ | $\sigma_m$ | $\sigma_e$ | $\sigma_o$ | $\sigma_{f_U}$ | $\sigma_{f_H}$ |
| 3% | 20% | 20% | 3% | 3% | 4% | 4% |

Figure 3 displays the expected precision in the geo-neutrino rates and thorium to uranium ratio of the reference signal for each detector as a function of observation time. Background from reactor antineutrinos significantly degrades the precision. As expected, continental detectors most precisely measure the background-subtracted crust rate, while the oceanic detector most precisely measures the background-subtracted mantle rate. Larger detectors reach the limit of systematic uncertainty more quickly than the smaller detectors. Table 7 presents the observation time required for the statistical uncertainty to equal the systematic uncertainty for each measurement.

### Discussion

Resolution of geological models is very unlikely with the currently operating geo-neutrino detectors. The difficulty for KamLAND is the inflated systematic uncertainty due to the high reactor background at the Kamioka site, while statistical uncertainty due to slowly accruing exposure is the problem for Borexino. The detector under construction on the Canadian Shield, SNO+, could resolve crustal models if operated for more than five years without significant increases to the reactor background. Operation of the proposed, larger continental observatories, Baksan, DUSEL, and LENA, would resolve crustal models in a relatively short time. For example, the proposed Baksan detector, which is about one-tenth the size of the other proposed continental detectors, would resolve crustal models with about one year of operation. Of the continental locations, the DUSEL site is marginally superior due to the lower reactor background. Resolution of mantle models requires the operation of an oceanic observatory, such as Hanohano. Systematic uncertainty in the crustal rate limits the background-subtracted mantle rate measurement to about 100% at any continental observatory. This study finds that the proposed Hanohano observatory needs to operate for about five years in order to achieve 10% precision. A deployment in the deep ocean of this duration could be a challenging constraint. Estimating the thorium to uranium ratio with 20% precision is possible with both continental and oceanic observatories, thus providing assessments of both the crustal and mantle reservoirs. Reactor background significantly affects the precision of this measurement, severely limiting KamLAND and to a lesser extent SNO+.

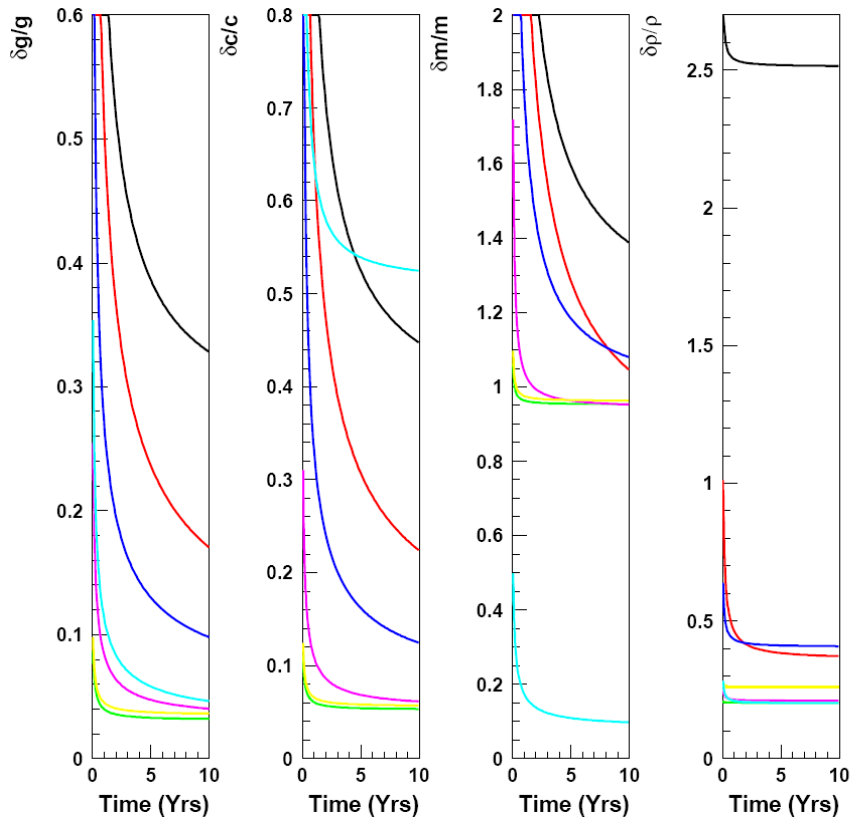

**Fig. 3.** Precision of the background-subtracted event rates of all geo-neutrinos, crustal geo-neutrinos, and mantle geo-neutrinos, and the ratio of geo-neutrino events from thorium and uranium as a function of observation time for the detectors in this study (Black=KamLAND; Red=Borexino; Blue=SNO+; Green=DUSEL; Purple=Baksan; Yellow=LENA; Turquoise=Hanohano).

**Table 7**
Listing for each detector of the exposure time in years required for the statistical uncertainty to equal the systematic uncertainty and the systematic uncertainty in percent.

|  | KamLAND | | Borexino | | SNO+ | | DUSEL | | Baksan | | LENA | | Hanohano | |
| --- | --- | --- | --- | --- | --- | --- | --- | --- | --- | --- | --- | --- | --- | --- |
|  | t | % | t | % | t | % | t | % | t | % | t | % | t | % |
| Geo-nu | 6.3 | 26 | 146 | 4.3 | 33 | 4.7 | 0.7 | 3.1 | 6.5 | 3.1 | 0.7 | 3.5 | 14 | 3.0 |
| Crust | 6.1 | 35 | 79 | 7.5 | 23 | 6.8 | 0.4 | 5.2 | 3.3 | 5.3 | 0.4 | 5.6 | 0.6 | 51 |
| Mantle | 4.8 | 114 | 11 | 73 | 2.6 | 96 | .02 | 95 | 0.2 | 94 | .03 | 96 | 3.4 | 8.4 |
| Th/U | 2.3 | 251 | 69 | 36 | 15 | 40 | 0.5 | 20 | 4.5 | 21 | 0.4 | 26 | 9.9 | 20 |

Relaxing the requirement of resolving individual geological models to that of constraining combinations of models is promising for SNO+. KamLAND reports a geo-neutrino flux of $(4.4\pm1.6) \times 10^6$ cm$^{-2}$ s$^{-1}$ measured with an exposure of $2.44 \times 10^{32}$ p$^+$ y (Abe et al., 2008), which is in agreement with a detailed prediction (Enomoto et al., 2007). Figure 4 compares this latest measurement with the geological models predicting the most and least intense fluxes from the crust and mantle. It is evident that the 36% precision of the current KamLAND measurement is insufficient to constrain combinations of geological models. A similar exposure of SNO+, however, would produce a measurement of the geo-neutrino flux with precision improved by about a factor of three. This is sufficient for constraining model combinations predicting the most and least intense geo-neutrino fluxes.

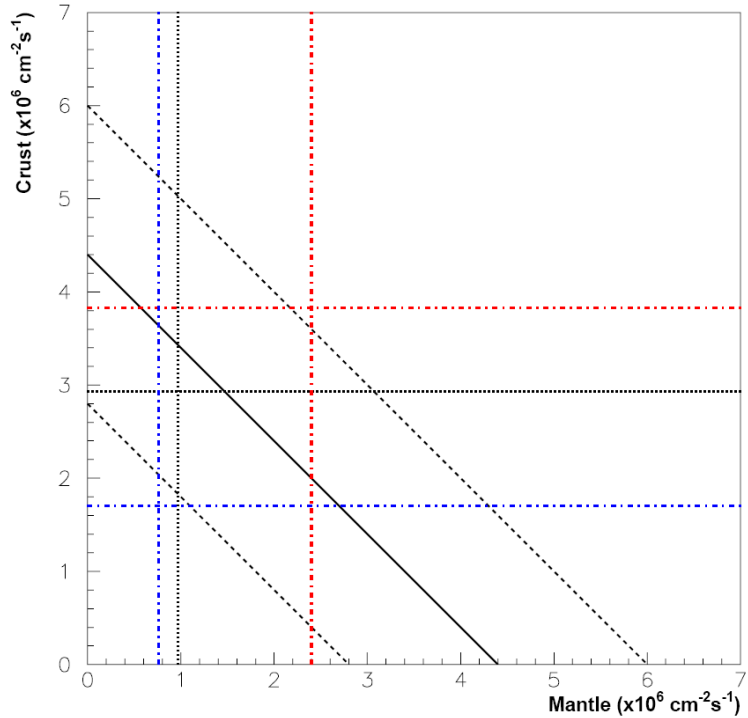

**Fig. 4.** The geo-neutrino flux measurement by KamLAND (solid black line) plots diagonally with the dashed lines showing the uncertainty. For comparison geological models predicting the most intense fluxes (red lines) and the least intense fluxes (blue lines) plot horizontally for the crust and vertically for the mantle. The reference signal (black dots) is in agreement with the KamLAND measurement.

## Conclusions

Geo-neutrino measurements are sensitive to the quantities and distributions of terrestrial uranium and thorium. The current KamLAND and Borexino projects are measuring geo-neutrinos. SNO+, which is under construction, can contribute to the resolution of crustal models after about five years of operation and exclude combinations of crustal and mantle models predicting maximum or minimum geo-neutrino fluxes. Together these projects are providing the experience necessary to complete larger-scale, proposed projects. Geo-neutrino measurements by proposed continental projects, Baksan, LENA, and DUSEL, are capable of resolving geological models of the continental crust with about one year of operation. An oceanic project the size of Hanohano effectively resolves mantle models with about five years of operation. The resolution of geological models of the continental crust and mantle would improve understanding of the origin of the earth and its thermal history by constraining its composition and radiogenic power.

## Acknowledgements

The author is grateful to John Mahoney for comments.